\documentstyle[preprint,aps,epsf]{revtex}

\def\eqn#1{Eq.~(\ref{#1})}

\def\slash#1{\rlap{$#1$}/} 

\def\tbar{{  ${\bf\overline{3}}$ }}
\def\ss{{ {\bf 6} }}

\def\sigcpp{{\Sigma_c^{++}}}
\def\sigcp{{\Sigma_c^{+}}}
\def\sigcz{{  \Sigma_c^{0}  }}
\def\xiccz{{  \Xi_{c2}^{0}  }}
\def\xiccp{{  \Xi_{c2}^{+}  }}
\def\omc{{  \Omega_c^{0}  }}
\def\xiczs{{  \Xi_{c2}^{0*}  }}

\begin{document}

{\tighten
\preprint{\vbox{
\hbox{CMU-HEP95-11}
\hbox{DOE-ER/40682-101}
\hbox{13A-40561-INT95-017}
\hbox{DOE Research and }
\hbox{Development Report}
}}

\title{Charmed Baryon Masses in Chiral Perturbation Theory
 \footnote{Work supported in part by
the U.S. Dept. of Energy under Grant No. DE-FG02-91-ER40682}}

\author{Martin J. Savage\footnote{DOE Outstanding Junior Investigator}}
\address{Department of Physics, Carnegie Mellon University\\
Pittsburgh, Pennsylvania 15213
U.S.A.\\ {\tt savage@thepub.phys.cmu.edu}}

\bigskip
\date{August 1995}

\maketitle
\begin{abstract}

The masses of the charmed baryons in the ${\bf 6}$ representation
of $SU(3)$ obey an equal spacing rule at lowest order in
$SU(3)$ breaking, ${\cal O}(m_s)$.
We compute the corrections to this relation at order
${\cal O}(m_s^{3/2})$
arising from meson loops using chiral perturbation theory
combined with heavy quark symmetry and find them to be small.
We also examine the hyperfine interaction responsible for the
splitting between the $J^\pi={3\over 2}^+$ and $J^\pi={1\over 2}^+$
baryons in the \ss\ representation.
The results also hold in the b-baryon sector.

\end{abstract}
\vfill\eject

Our knowledge of the masses and properties of the lowest lying charmed baryons
has improved
dramatically during the last few years \cite{PDG94}.
With the recent discovery of the $\xiczs$ ($J^\pi = {3\over 2}^+$) of mass
$2642.8\pm 2.2 \ {\rm MeV}$ \cite{CLEO95a}\
and hints of the $\xiccp$ ($J^\pi = {1\over 2}^+$) with a mass of
$\sim 2560\  {\rm MeV}$ \cite{WA8995a}\
and $\Sigma_c^{++*}$ ($J^\pi = {3\over 2}^+$) with a mass of
$\sim 2530\  {\rm MeV}$
\cite{SKAT93} it has become timely to see just how well we understand
the pattern of masses
in the charmed baryon sector.
There have been many estimates of the charmed baryon masses made in the
past \cite{previous}.
Recently, Rosner  \cite{Ros95a}\ has performed a
spin-flavour analysis of the
charmed baryons and the lowest lying noncharmed baryons to obtain
masses and strong decay widths.
He obtained mass relations between the $J^\pi = {3\over 2}^+$ and
$J^\pi = {1\over 2}^+$ charmed baryons
in the lowest lying \ss\ representations including SU(3) breaking
arising from the
difference between the strange and non-strange constituent quark masses.
In this work we will examine SU(3) breaking in the lowest lying \ss\
representation of charmed baryons
using chiral perturbation theory with heavy quark symmetry.  At lowest
order in SU(3) breaking there is an equal-spacing rule
analogous to the equal spacing rule in the noncharmed baryon decuplet
that receives finite and computable corrections.
We determine that these corrections are small.
The hyperfine mass splittings between the \ss\ and $\ss^*$
are also examined.

Heavy quark symmetry and chiral symmetry are combined together
in order to describe the soft hadronic interactions of hadrons containing a
heavy quark \cite{HCPT,Cho92a}.
The light degrees of freedom in the ground state of a baryon containing one
heavy
quark can have $s_l=0$ corresponding to a member of the
flavour $SU(3)$ \tbar\  , $T_i (v)$,  with $J^\pi = {1\over 2}^+$
or they can have $s_l=1$ corresponding to a member of the flavour $SU(3)$ \ss\
,
$S^{ij}_\mu (v)$.
In the latter case, the spin of the light degrees of freedom can be combined
with the spin of the heavy
quark to form both $J^\pi={3\over 2}^+$ and $J^\pi={1\over 2}^+$ baryons, which
are
degenerate in the $m_Q\rightarrow \infty$ limit.
Baryons in the \tbar\ and \ss\ representations are described by the fields
\begin{eqnarray}
& S^{ij}_\mu (v) =  {1\over\sqrt{3}} (\gamma_\mu + v_\mu) \gamma_5
{1\over 2} (1+\slash{v}) B^{ij}  \ + \ {1\over 2}(1+\slash{v}) B^{*ij}_\mu
\nonumber\\
& T_i (v) =  {1\over 2}(1+\slash{v}) B_i
\ \ \ \ ,
\end{eqnarray}
where the $J^\pi={1\over 2}^+$ charmed baryons of the \ss\ are assigned to
the symmetric tensor $B^{ij}$
\begin{eqnarray}
B^{11} & = \Sigma_c^{++}\ ,\ B^{12} = {1\over\sqrt{2}}\Sigma_c^+\ ,\ B^{22} =
\Sigma_c^0 \ ,\nonumber\\
B^{13} & = {1\over\sqrt{2}}\Xi_{c2}^+ \ ,\ B^{23} = {1\over\sqrt{2}}\Xi_{c2}^0
\
,\
B^{33} = \Omega_c^0 \ \ \ .
\end{eqnarray}
The $J^\pi={3\over 2}^+$ partners of these baryons have the same
$SU(3)$ assignment in $B^{*ij}_\mu$.
The charmed baryons of the \tbar\ representation are assigned to $B_i$ as
\begin{eqnarray}
B_1 = \Xi_{c1}^0 \  , \ B_2 = -\Xi_{c1}^+ \ , \ B_3 = \Lambda_c^+ \ \ \ .
\end{eqnarray}
The chiral lagrangian describing SU(3) invariant soft hadronic interactions of
the
charmed baryons is \cite{Cho92a}
\begin{eqnarray}\label{lagQ}
& {\cal L}_Q =   i\overline{T^i} v\cdot D T_i \ -\  i\overline{S}^\mu_{ij}
v\cdot D S_\mu^{ij} \ +\ \Delta_0 \overline{T^i} T_i
+ {f^2\over 8} Tr\left[ \partial^\mu\Sigma\partial_\mu\Sigma^\dagger \right]
\nonumber\\
&\ + \ g_3\left(  \epsilon_{ijk}\overline{T}^i (A^\mu)^j_l S^{kl}_\mu + h.c.
\right)
+  \ i g_2\ \epsilon_{\mu\nu\rho\sigma}\overline{S}^\mu_{ik} v^\nu(A^\rho)^i_j
S^{\sigma jk} \ +\ \cdots
\ \ \ ,
\end{eqnarray}
where the dots denote operators with
more derivatives or those that are
higher order in the $1/m_Q$ expansion and
$D^\alpha$ is the chiral covariant derivative.
The axial chiral field
$A^\mu=~{i \over 2}\left( \xi^\dagger\partial^\mu\xi -
\xi\partial^\mu\xi^\dagger \right) $
is defined in terms of
$\xi = {\rm exp}\left(iM/f\right)$
where $M$ is the octet of pseudo-Goldstone bosons
\begin{eqnarray}
 M = \left(
\matrix{ {1\over\sqrt{6}}\eta + {1\over\sqrt{2}}\pi^0 & \pi^+ & K^+ \cr
\pi^- & {1\over\sqrt{6}}\eta - {1\over\sqrt{2}}\pi^0 & K^0 \cr
K^- & \overline{K}^0 & -{2\over\sqrt{6}}\eta \cr } \right)
\ \ \ \ \ ,
\end{eqnarray}
and $f \sim 132 {\rm MeV}$ is the pion decay constant at lowest order.
The $\Sigma$ field of pseudo-Goldstone bosons is
$\Sigma = \xi^2 = {\rm exp}\left(i 2 M/f\right)$.
Coupling of a single pseudo-Goldstone boson to the \tbar\  baryons
is forbidden at lowest order in $1/m_Q$.
Even in the infinite mass limit the \ss\ baryons are not degenerate
with the \tbar\  baryons as the light degrees of freedom are
in a different configuration giving rise to an intrinsic mass difference
$\Delta_0$.  We have chosen to remove the mass of the \ss\ from the fields
and not the mass of the \tbar\  for convenience.
The masses of the charmed baryons that follow from \eqn{lagQ} are trivial in
the sense that there is no SU(3) breaking
and the charmed baryons in the \tbar have equal mass, as do those
in the \ss\ but the \tbar and \ss\ are split by  $\Delta_0$.

The strong coupling constants $g_2$ and $g_3$ must be determined from
experimental data on the
strong widths or from loop processes.  Observation of the $\xiczs$ and
the upper limit on its width
\cite{CLEO95a}\ of $\Gamma ( \xiczs) < 5.5 \ {\rm MeV}$ constrains $g_3$
(neglecting higher order corrections) to be
$|g_3| < 1.3$.    The coupling constant $g_2$ is, as yet, unconstrained.
We notice that the upper bound on $g_3$ is already below the value one
would expect from large $N_c$ considerations \cite{GLM93a,Jen93a},
$g_2 = -{3\over 2} g_A = -1.9$ and $g_3 = \sqrt{3\over 2} g_A = 1.5$
with $g_A\sim 1.25$.

SU(3) breaking in the masses of the charmed baryons arises from explicit
insertions
of the light quark mass matrix and from loop graphs involving the
pseudoscalar  mesons.
The general form of such corrections is discussed in \cite{CCLLYY94a}\ ;
however,  we wish to be more specific.
Using the notation of \cite{CCLLYY94a} we write the lagrange density
that is
linear in the light quark mass matrix  and is lowest order in the heavy
 quark expansion
\begin{eqnarray}\label{lagbreak}
{\cal L} & =
\lambda_1 \overline{S}^\mu_{ij} (\chi^+)^i_k S_\mu^{jk}
+ \lambda_2 \overline{S}^\mu_{ij} S_\mu^{ij} (\chi^+)^k_k
+ \lambda_3 \overline{T^i} (\chi^+)_i^j T_j
+\lambda_4 \overline{T^i} T_i (\chi^+)^k_k
 - \mu (\chi^+)^k_k
\ \ \ \ ,
\end{eqnarray}
where
\begin{eqnarray}
\chi^+ = \xi^\dagger m_q \xi^\dagger + \xi m_q \xi
\ \ \ ,
\end{eqnarray}
and where the light quark mass matrix is
\begin{eqnarray}
m_q = \left(\matrix{m_u&0&0\cr 0&m_d&0\cr 0&0&m_s}\right)
\ \ \  .
\end{eqnarray}
The last term in \eqn{lagbreak}\  generates the non-zero masses of the
pseudo-Goldstone bosons while the
first four terms contribute to the masses of the charmed baryons.
Non-zero values for
$\lambda_1$ and $\lambda_3$ give rise to the leading SU(3) breaking
between charmed baryon masses.
Performing the contraction of indices in \eqn{lagbreak} one generates
baryon masses in the presence of octet $SU(3)$ breaking.
In general, the \ss\  could have breaking terms with representations
${\bf 6} \otimes\overline{\bf 6} =
{\bf 27}\oplus {\bf 8} \oplus {\bf 1}$ but the single insertion of
$m_q$ gives octet breaking only.
As there are only three elements in the \tbar\ and two are degenerate
in the limit of isospin symmetry
there are no $SU(3)$ mass relations that hold in the presence of
\eqn{lagbreak}.
However, there is a nontrivial mass relation between baryons in the
\ss\
that holds in the presence of \eqn{lagbreak},
\begin{eqnarray}\label{mesr}
{1\over 3}\left( M_\sigcpp + M_\sigcp + M_\sigcz \right) + M_\omc
- \left( M_\xiccp + M_\xiccz \right) = 0
\ \ \ ,
\end{eqnarray}
which in the limit of isospin symmetry becomes
\begin{eqnarray}\label{esrule}
M_{\Sigma_c}  + M_{\Omega_c} - 2 M_{\Xi_{c2}} = 0
 \ \ \ .
\end{eqnarray}
This is an equal spacing rule analogous to the equal spacing rule
that arises in the
decuplet of uncharmed
$J^\pi = {3\over 2}^+$ baryons.
We have not yet discussed mixing between the $\Xi_c$'s in the \tbar\
and \ss\ .
Mixing between these particles is both SU(3) breaking and a $1/M_Q$
effect
(as it requires mixing between states with $s_l=0$ and $s_l=1$ in
the heavy quark limit).
Further, the mixing term will enter squared when the
$ \{\Xi_{c1}, \Xi_{c2} \}$  mass matrix is
diagonalized and therefore we neglect it.

Corrections to the equal spacing rule \eqn{esrule} arise from more
insertions of the
light quark mass matrix and from loops involving the
pseudo-Goldstone bosons.
The leading corrections arise from the meson loops and are of order
$m_s^{3/2}$.
Graphs involving the $\ss^{(*)}$  baryons depend on the meson masses
only (neglecting the ${\cal O}(m_s)$ splittings between the intermediate
state baryons
within the loop that give corrections higher order in $m_s$),
while those involving the
\tbar\ baryons are functions of the meson masses and the
mass splitting $\Delta_0$.
Despite the loop contribution to the individual masses being of order a
few hundred MeV
(expressions for which can be derived from results in \cite{CCLLYY94a}),
the correction to the equal spacing rule is small; explicitly we find that
\begin{eqnarray}\label{esrulecorr}
M_{\Sigma_c}  + M_{\Omega_c} - 2 M_{\Xi_c} =
{1\over 48\pi f^2}\left[  g_2^2 {\cal J}(0) - g_3^2 {\cal J}(\Delta_0)
\right]
 \ \ \ ,
\end{eqnarray}
where the function ${\cal J}(y)$ is given by
\begin{eqnarray}
& {\cal J}(y) = {1\over \pi} \bigg[
-y^3 \log\left(m_K^8 / m_\eta^6 m_\pi^2\right)
+ 6y m_K^2\log\left(m_K^2/ m_\pi^2\right)
-{9\over 2} y m_\eta^2\log \left(m_\eta^2/ m_\pi^2\right)
\nonumber\\
&  - 4 {\cal G}(y,m_K) + 3 {\cal G}(y,m_\eta) + {\cal G}(y,m_\pi)
\bigg]
\end{eqnarray}
and
\begin{eqnarray}
{\cal G}(y,m) = (y^2-m^2)^{3\over 2}
\log\left({-y-\sqrt{y^2-m^2+ i \epsilon}\over -y+\sqrt{y^2-m^2+i\epsilon}
}\right)
\ \ \ \ .
\end{eqnarray}
In order to arrive at this result we have used the Gell-Mann--Okubo mass
relation between the pseudo-Goldstone boson masses
$4m_k^2-3 m_\eta^2-m_\pi^2 = 0$
that arises at lowest order from \eqn{lagQ} and \eqn{lagbreak}.
Notice that the correction to the equal-spacing rule
\eqn{esrule}\ (and \eqn{mesr})
that holds for ${\bf 8}\oplus {\bf 1}$ $SU(3)$ breaking is finite.
This results from the fact that any corrections to
\eqn{esrule}\ and \eqn{mesr} must transform as  a  ${\bf 27}$ under
$SU(3)$ and there are no counterterms in \eqn{lagQ}\
and \eqn{lagbreak}\  that transform as a ${\bf 27}$ to absorb divergences
(such counterterms start at ${\cal O}(m_s^2)$).
In the limit of vanishing \ss-\tbar mass splitting $\Delta_0$ we have
\begin{eqnarray}\label{twentyseven}
{\cal J}(0) = 4 m_K^3 - 3 m_\eta^3 - m_\pi^3
\ \ \ \ .
\end{eqnarray}
This particular combination of masses appears in the violation of
mass relations that hold in the presence of
octet $SU(3)$ breaking in other hadronic sectors,
in the octet baryon sector \cite{Jen92a}\
and in the vector meson sector \cite{JMW95a}.
This is because a ${\bf 27}$ representation is required to violate the
mass relations in each sector
and the combination of masses that appears in \eqn{twentyseven}\
is the unique combination that transforms as a ${\bf 27}$
\cite{JMW95a,Jenprep}.
Numerically,  we find that the right hand side of \eqn{esrulecorr}\ is
very small
$\sim 5 (g_2^2-g_3^2) {\rm MeV}$ using the masses of the charged $K$ and
$\pi$ and
setting $\Delta_0=100\  {\rm MeV}$ (the result is very insensitive to the
value of $\Delta_0$).
Therefore, we expect that the equal-spacing rule in \eqn{esrule}\ is well
satisfied.
We can use the mass of the $\Sigma_c^{++}$ , $2453.1\pm 0.6\  {\rm MeV}$ and
the mass of the
$\Omega_c^0$, $2704\pm 4 \ {\rm MeV}$ to predict that
\begin{eqnarray}\label{massnum}
M_{\Xi_{c2}} =  &  {1\over 2} \left[ M_{\Sigma_c} + M_{\Omega_c} \right]
\nonumber\\
 \sim & 2579 \ {\rm MeV}
\ \ \ \ ,
\end{eqnarray}
which  we expect to be within a few ${\rm MeV}$ of the actual mass.
This is in contrast to the recent experimental suggestion \cite{WA8995a}\
of
$M_{\Xi_{c2}} \sim 2560 \ {\rm MeV}$, some $20\  {\rm MeV}$ away from
\eqn{massnum}.
The mixing between the \ss\  and \tbar\  that we have neglected in
our analysis will only
increase the mass computed in \eqn{massnum}, further increasing
the possible discrepancy.
It seems best not to consider this a serious problem or to compute
higher order corrections to
\eqn{esrule}\ until the experimental situation becomes more certain.

Turning now to the hyperfine mass splitting between the \ss\  and $\ss^*$.
Such a splitting results from the charm quark not being infinitely more
massive than the scale of strong interactions.
One can make a crude estimate for the magnitude of the splitting of
$\delta_6 \sim \Lambda_{\rm QCD}^2/m_c \sim \ 50 \ {\rm MeV}$.
Using the $\xiczs$ mass measurement \cite{CLEO95a}
and the $\Xi_{c2}$ mass determined from the
\eqn{massnum}\ we find $\delta_6 \sim\  64\  {\rm MeV}$,  consistent with our
naive estimate.
The hyperfine mass splittings are induced at lowest order by the $SU(3)$
invariant
lagrange density \cite{MJS95a}
\begin{eqnarray}
{\cal L} = {\delta_6\over 6} ( g_{\mu\alpha}g_{\nu\beta} -
g_{\nu\alpha}g_{\mu\beta} )
\overline{S}^\mu_{ij} i \sigma^{\alpha\beta} S^{\nu ij}
\ \ \ .
\end{eqnarray}
It is clear that this operator gives rise to an equal-spacing rule for the
hyperfine mass splittings
( we use $\delta_{\Sigma_c} = M_{\Sigma_c^*}-M_{\Sigma_c} $ ,
$\delta_{\Xi_c} = M_{\Xi_c^*}-M_{\Xi_c}$
and $\delta_{\Omega_c} = M_{\Omega_c^*}-M_{\Omega_c}$ for compactness)
\begin{eqnarray}
\delta_{\Sigma_c} = \delta_{\Xi_c}  = \delta_{\Omega_c}
\ \ \ ,
\end{eqnarray}
from which we predict that
\begin{eqnarray}\label{hyplow}
M_{\Sigma_c^*} \sim 2518 \ {\rm MeV}
\ \ \ {\rm and }\ \ \
M_{\Omega_c^*} \sim 2768 \ {\rm MeV}
\ \ \ \ .
\end{eqnarray}
The values in \eqn{hyplow}\  are both within $4 \ {\rm MeV}$ of the masses
predicted by Rosner \cite{Ros95a}\
using spin-flavour wavefunctions.
Note that we have used the hyperfine mass splitting based
on the equal-spacing rule prediction for
the mass of the $\Xi_{c2}$.  If we had used the value suggested by
\cite{WA8995a}\  then the predicted masses of the
$\Sigma_c^*$ and $\Omega_c^*$ would be $\sim 2538 \ {\rm MeV}$ and
$\sim 2788 \ {\rm MeV}$ respectively.

When we consider $SU(3)$ breaking of the hyperfine mass splittings, an
equal spacing rule analogous to that in
\eqn{esrule}\ follows at linear order in $m_s$ (assuming isospin symmetry),
\begin{eqnarray}\label{hypesr}
\delta_{\Sigma_c} + \delta_{\Omega_c}  - 2 \delta_{\Xi_c}  = 0
\ \ \ ,
\end{eqnarray}
from the lagrange density
\begin{eqnarray}
{\cal L} = {\delta_6^\prime\over 6}
( g_{\mu\alpha}g_{\nu\beta} - g_{\nu\alpha}g_{\mu\beta} )
\overline{S}^\mu_{ij} i \sigma^{\alpha\beta} S^{\nu ik} (\chi^+)_k^j
+
 {\delta_6^{\prime\prime}\over 6}
( g_{\mu\alpha}g_{\nu\beta} - g_{\nu\alpha}g_{\mu\beta} )
\overline{S}^\mu_{ij} i \sigma^{\alpha\beta} S^{\nu ij} (\chi^+)_k^k
\ \ \ \ .
\end{eqnarray}
As the hyperfine mass splittings are a $1/m_c$ effect the leading
$SU(3)$ breaking corrections to these splittings from meson loops
must vanish as  $\delta_6$ vanishes.
This means that the leading corrections to \eqn{hypesr}\ are not $m_s^{3/2}$
but  $m_s \log m_s$,
when $\delta_6$ is treated as small.
Explicit computation of the pseudo-Goldstone boson loop graphs
(setting $\Delta_0=0$)
gives a finite correction to \eqn{hypesr} of
\begin{eqnarray}\label{hypcorr}
\delta_{\Sigma_c} + \delta_{\Omega_c}  - 2 \delta_{\Xi_c} =
{\delta_6 (3 g_2^2-2 g_3^2) \over 16\pi^2 f^2}
\left[  m_k^2 \log\left( {m_K^2\over m_\pi^2} \right) -
{3\over 4} m_\eta^2 \log \left({ m_\eta^2\over m_\pi^2} \right)
\right]
\ \ \ ,
\end{eqnarray}
where we have again used the Gell-Mann--Okubo mass formula for the
mesons to simplify the expression.
We see that the correction to the hyperfine equal-spacing rule in
\eqn{hypcorr}\
is small ($\sim 10^{-2}\delta_6 (3 g_2^2-2 g_3^2)$)
and therefore we expect the equal-spacing rule to be well satisfied.

Unlike the $SU(3)$ corrections to the individual baryon masses from
meson loops in the heavy quark limit
which are finite when $\Delta_0\rightarrow 0$, the loop corrections to
the individual hyperfine splittings are divergent,
and require presently unknown counterterms to absorb the divergence.
The nonanalytic contributions from the loop graphs are of the form
$m_M^2\log\left( m_M/\Lambda_\chi \right)$
where we have renormalized at the chiral symmetry breaking scale
 $\Lambda_\chi$.
The same type of corrections contribute to the hyperfine mass
splittings between the vector and
pseudoscalar mesons containing a heavy quark.
It is found that the
counterterm required to reproduce the observed spectrum
essentially exactly cancels the contribution of the chiral
logarithm \cite{Jen94a}.
This leads one to believe that chiral perturbation theory, in
 particular the neglect of the counterterms,
may be failing for the hyperfine  mass splittings.
However, as the hyperfine equal-spacing rule in \eqn{hypcorr}\
is independent of the
counterterms, one might hope that the loop graphs do give a
reasonable estimate of the size of the violation.

In conclusion, there is an equal-spacing rule that holds for the masses
of the charmed baryons in the
\ss\ representation of $SU(3)$ in the presence of octet
$SU(3)$ breaking.
Violations of this equal-spacing rule arise  at leading order from
loop graphs involving the pseudo-Goldstone bosons and
behave as $m_s^{3/2}$.
As the violation must transform as a ${\bf 27}$ under $SU(3)$ we find
that the combination of meson
masses that enters is numerically very small and we expect the
equal-spacing rule to be well satisfied.
There is also an equal-spacing rule for the hyperfine mass splittings
between the charmed baryons in the
\ss\  and the $\ss^*$.
This also receives corrections from meson loop graphs but of the
form $m_s\log m_s$
(a consequence of heavy quark symmetry).
Despite the contribution to individual hyperfine splittings being
 divergent and
requiring the presence of an unknown counterterm
the equal spacing rule that holds in the presence of
octet
$SU(3)$ breaking receives only a finite and  numerically small
correction.
Therefore, we expect that the equal spacing rules that
hold in the presence of octet $SU(3)$ breaking ,
$M_{\Sigma_c} + M_{\Omega_c} - 2 M_{\Xi_c} = 0$ and
$\delta_{\Sigma_c} + \delta_{\Omega_c} - 2 \delta_{\Xi_c} = 0$,
are well satisfied in nature.
This is not because the loop corrections to the baryons
masses are small in chiral perturbation theory (they are not)
but because the group structure forces any violation of
relations that hold in the presence of
octet $SU(3)$ breaking to be small.

The equal-spacing rules and their loop corrections
in the charmed baryon sector have direct analogues in the b-baryon sector.
We have that
$M_{\Sigma_b} + M_{\Omega_b} - 2 M_{\Xi_b} = 0$ and
$\delta_{\Sigma_b}~+~\delta_{\Omega_b}~-~2 \delta_{\Xi_b}~=~0$,
with only small corrections from meson loops.

\bigskip

\centerline{\bf Acknowledgements}

I would like to thank Jonathan Rosner for stimulating this work
and for invaluable discussions and
Elizabeth Jenkins for an important discussion.
I would also like to thank the Institute for Nuclear theory at the
University of Washington
for kind hospitality during the course of this work.

\end{document}